\documentclass[journal=jacsat,manuscript=article]{achemso}

\usepackage[version=3]{mhchem}
\usepackage{dutchcal}
\usepackage{hyperref}
\usepackage{multirow}
\usepackage{dcolumn}
\usepackage{cellspace}
\usepackage{longtable}
\usepackage{dcolumn}
\usepackage{bm}
\usepackage{array}
\usepackage{pdfpages}

\renewcommand{\exp}[1]{\text{exp}({#1})}

\renewcommand{\mathbf}[1]{\bm{#1}}

\newcommand{\ketT}[1]{|\text{#1}\rangle}
\newcommand{\braT}[1]{\langle\text{#1}|}
\newcommand{\ketM}[1]{|#1\rangle}
\newcommand{\braM}[1]{\langle#1|}
\newcommand{\braMketM}[2]{\langle#1|#2\rangle}

\author{Leo Stoll}
\author{Sara Angelico}
\author{Eirik F. Kjønstad}
\author{Henrik Koch}
\affiliation[NTNU]{Department of Chemistry, Norwegian University of Science and Technology, Trondheim, Norway}
\email{henrik.koch@ntnu.no}

\title[]{Similarity Constrained CC2 for Efficient Coupled Cluster 
Nonadiabatic Dynamics}

\begin{document}

\begin{abstract}
    Despite their high accuracy, standard coupled cluster models cannot be used for nonadiabatic molecular dynamics simulations because they yield unphysical complex excitation energies at conical intersections between same-symmetry excited states. On the other hand, similarity constrained coupled cluster theory has enabled the application of coupled cluster theory in such dynamics simulations. Here, we present a similarity constrained perturbative doubles (SCC2) model with same-symmetry excited-state conical intersections that exhibit correct topography, topology, and real excitation energies. This is achieved while retaining the favorable computational scaling of the standard CC2 model. We illustrate the model for conical intersections in hypofluorous acid and thymine, and compare its performance with other methods. The results demonstrate that conical intersections between excited states can be described correctly and efficiently at the SCC2 level. We therefore expect that the SCC2 model will enable coupled cluster nonadiabatic dynamics simulations for large molecular systems.\\
\end{abstract}

\section{Introduction}
Conical intersections have been shown to play a pivotal role in the excited state dynamics of most molecular systems, both through theoretical studies \cite{NonadiabaticityTheImportanceOfConicalIntersections, NonadiabaticEventsandConicalIntersections} and experiments using pump-probe techniques.\cite{FemtosecondTimeResolvedPhotoelectronSpectroscopyofPolyatomicMolecules} These intersections facilitate internal conversion as an ultra-fast relaxation process, typically occurring tens to hundreds of femtoseconds after excitation of the system.\cite{domcke2004conical} When a molecular system approaches a conical intersection, the nuclear and electronic motions become coupled, and the Born-Oppenheimer approximation breaks down. This necessitates the use of nonadiabatic dynamics simulation methods to model the time evolution of the system.\cite{AbInitioNonadiabaticQuantumMolecularDynamics} Such methods rely on a balanced treatment of all electronic states involved in the dynamics, and are sensitive to the accuracy of the applied electronic structure model.\cite{MolecularPhotochemistryRecentDevelopmentsinTheory}

Coupled cluster (CC) methods are among the most accurate electronic structure methods broadly available in quantum chemistry software.\cite{Coupledclustertheoryinquantumchemistry} Despite their generally steep computational scaling, they have found a wide range of applications where a highly accurate description of the electronic structure is required. In excited states, both static and dynamic correlation can be treated effectively through equation of motion (EOM-CC) \cite{TheEquationOfMotionCoupledClusterMethod} or linear response \cite{Coupledclusterresponsefunctions} coupled cluster theory. Within the coupled cluster hierarchy, the second order approximate coupled cluster singles and doubles model (CC2) \cite{TheSecondOrderApproximateCC2Model} provides a unique trade-off between comparatively fast computation times and treatment of electron correlation.\cite{AMountaineeringStrategytoExcitedStates:HighlyAccurateEnergiesandBenchmarksforMediumSizedMolecules} For systems dominated by a single excited configuration, the CC2 model typically provides equilibrium- and excited state energies with errors of few tenths of an eV.\cite{AMountaineeringStrategytoExcitedStates:HighlyAccurateEnergiesandBenchmarksforMediumSizedMolecules,BenchmarksForElectronicallyExcitedStates:CASPT2CC2CCSDCC3} When using medium-sized basis-sets, CC2 may be applied to systems of hundreds of atoms in single-point calculations. This balance between accuracy and low computational cost makes the CC2 model an especially attractive choice for coupled cluster nonadiabatic dynamics simulations.

However, already two decades ago, Hättig \cite{StructureOptimizationForExcitedStatesWithCorrelatedSecondORderMethodsCC2ADC(2)} pointed out that the non-Hermitian effective Hamiltonian of coupled cluster theory may produce unphysical artifacts (such as complex energies) in the description of conical intersections between excited states. For calculations at equilibrium geometries, such artifacts are typically avoided, but they are rapidly encountered when the excited state potential energy surfaces are explored
in nonadiabatic dynamics simulations. The appearance of complex excitation energies in close proximity to conical intersections between excited states of the same symmetry has been documented\cite{Cancoupledclustertheorytreatconicalintersections, CrossingConditionsInCoupledClusterTheory, unexpectedhydrogendissociationthymine, SurfaceHoppingDynamicswithCorrelatedSingleReferenceMethods9HAdenine, QuantumChemicalInvestigationoftheStructuresandElectronicSpectraoftheNucleicAcidBasesattheCoupledClusterCC2Level} not only for CC2, but also for the coupled cluster singles and doubles (CCSD) \cite{AFullCCSDModel} and the coupled cluster singles, doubles and triples (CCSDT) \cite{ThefullCCSDTmodelformolecularelectronicstructure, ErratumThefullCCSDTmodelformolecularelectronicstructure} models. Because of the abundance of conical intersections and their key role in excited state dynamics, the flawed description provided by standard coupled cluster methods has historically hindered their successful application in nonadiabatic dynamics simulations.\cite{StructureOptimizationForExcitedStatesWithCorrelatedSecondORderMethodsCC2ADC(2)} Specifically, the CC2 model failed in dynamics simulations on adenine due to the appearance of complex excitation energies.\cite{SurfaceHoppingDynamicswithCorrelatedSingleReferenceMethods9HAdenine}

Over the last decade, the development of similarity constrained coupled cluster (SCC)\cite{ResolvingTheNotoriousCaseOfConicalIntersectionsForCoupledClusterDynamics} theory has addressed the issues of standard coupled cluster models at conical intersections. These efforts recently led to the successful application of the similarity constrained coupled cluster singles and doubles model (SCCSD) \cite{AnOrbitalInvariantSimilarityConstrainedCoupledClusterModel} in nonadiabatic dynamics simulations on thymine, where standard CCSD encounters unphysical artifacts.\cite{unexpectedhydrogendissociationthymine} Nonetheless, the steep computational cost of the SCCSD model, which scales as $O(N^6)$ with the number of molecular orbitals (MOs) $N$, limits the size of system that can be studied. Therefore, in this work, we present a similarity constrained coupled cluster method which scales as $O(N^5)$. This SCC2 method maintains the accuracy of CC2 and is expected to enable coupled cluster nonadiabatic dynamics simulations for larger systems.

\section{Theory}\label{sec:Theory}

The coupled cluster wave function is defined as
\begin{equation}
    \ketT{CC} = \exp{T} \ketT{HF},
\end{equation}
where the cluster operator,
\begin{equation}\label{eq:clusteroperator}
    T = \sum_{\mu>0} t_\mu \tau_\mu,
\end{equation}
is a sum of excitation operators $\tau_\mu$ weighted by the cluster amplitudes $t_\mu$. The excitation operators act on the reference state, usually the Hartree-Fock state $\ketT{HF}$, to produce excited configurations $\ketM{\mu}$. We let $\mu=0$ refer to the reference state, corresponding to $\tau_0 = 1$ and $\ketM{0} = \ketT{HF}$. Similarly, $\mu_1$ and $\mu_2$ refer to single and double excitations from the reference state. Here we will restrict ourselves to singlet spin-adapted coupled cluster models, where the $\tau_\mu$ are restricted to singlet excitations.\cite{MolecularElectronicStructureTheory}

With the similarity transformed Hamiltonian defined as
\begin{equation} \label{eq:SimTransHamiltonian}
    \bar{H} = \exp{-T} H \exp{T},
\end{equation}
the cluster amplitudes are determined by the amplitude equations
\begin{equation} \label{eq:AmplitudeEquations}
    \Omega_\mu = \braM{\mu} \bar{H} \ketT{HF} = 0 \hspace{20pt} \mu > 0,
\end{equation}
and the coupled cluster energy is determined as
\begin{equation}\label{eq:GSEnergy}
    E_0 = \braT{HF} \bar{H} \ketT{HF}.
\end{equation}

The excited states can be obtained by equation of motion coupled cluster (EOM-CC)\cite{TheEquationOfMotionCoupledClusterMethod} theory. Right and left excited states, $\ketM{R^k}$ and $\braM{L^k}$, are expanded as
\begin{equation}\label{eq:EOM-CC}
\begin{aligned}
    \ketM{R^k} &= \sum_{\mu \geq 0} \exp{T} \ketM{\mu} r_\mu^k \\
    \braM{L^k} &= \sum_{\mu \geq 0} \braM{\mu} \exp{-T} l_\mu^k
\end{aligned}
\end{equation}
and are required to be biorthonormal, that is, $\braMketM{L^k}{R^l} = \delta_{kl}$.

The left and right EOM-CC expansion coefficients, $l_\mu^k$ and $r_\mu^k$, referred to as the excited state amplitudes, and the excitation energies $\omega_k$, are determined by the non-symmetric eigenvalue problem\citep{MolecularElectronicStructureTheory}
\begin{equation} \label{eq:UnsymmetricalEOMCC}
\begin{aligned}
    \mathbf{A} \mathbf{r}^k &= \omega_k \mathbf{r}^k \\
    \mathbf{A}^T \mathbf{l}^k&= \omega_k \mathbf{l}^k,
\end{aligned}
\end{equation}
where $\mathbf{l}^k$ and $\mathbf{r}^k$ exclude the ground state contributions $l^k_0$ and $r^k_0$.
The matrix $\mathbf{A}$ is the coupled cluster Jacobian, given by
\begin{equation} \label{eq:JacobianElements}
    A_{\mu\nu} = \braM{\mu} [\bar{H}, \tau_\nu] \ketT{HF} \hspace{15pt} \mu,\nu > 0.
\end{equation}

The coupled cluster Jacobian matrix enters the matrix representation of the similarity transformed Hamiltonian $\bar{H}$ in the basis $\{ \ketM{\mu} | \mu \geq 0 \}$,
\begin{equation}\label{eq:LevelShiftedHamiltonian}
    \mathbf{\bar{H}} = 
    \begin{pmatrix} 
        E_0 & \mathbf{\eta}^T \\
        \mathbf{0} & \mathbf{A} + E_0 \mathbf{I}
    \end{pmatrix},
\end{equation}
where the reference column (the first column) corresponds to the energy $E_0$ and amplitude equations $\mathbf{\Omega} = \mathbf{0}$, which are assumed to be solved. From the eigenvalue equation associated with this matrix, and the requirement of biorthogonality, the ground state contributions to the excited states can be determined as
\begin{equation}\label{eq:GroundStateContribution}
\begin{aligned}
    r^k_0 &= \frac{\mathbf{\eta}^T \mathbf{r}^k}{\omega_k}\\
    l^k_0 &= 0,
\end{aligned}
\end{equation}
where
\begin{equation} \label{eq:eta}
    \eta_\nu = \braT{HF} [\bar{H}, \tau_\nu] \ketT{HF}.
\end{equation}

By truncating the cluster operator in eqn (\ref{eq:clusteroperator}), different methods in the coupled cluster hierarchy can be obtained. In the following, we will restrict $T$ to include only single and double excitations, $T = T_1 + T_2$, where
\begin{equation}
    T_1 = \sum_{\mu_1} t_{\mu_1} \tau_{\mu_1} \hspace{20 pt} T_2 = \sum_{\mu_2} t_{\mu_2} \tau_{\mu_2}.
\end{equation}
This definition of $T$, together with eqs (\ref{eq:SimTransHamiltonian}), (\ref{eq:AmplitudeEquations}) and (\ref{eq:GSEnergy}), defines the CCSD model, which scales as $O(N^6)$.

\subsection{The CC2 model}
The electronic Hamiltonian can be divided into the Fock operator $F$ and the fluctuation potential $U$ as \cite{MolecularElectronicStructureTheory}
\begin{equation}
    H = F + U.
\end{equation}
The $O(N^5)$ scaling CC2 model is obtained from the CCSD model by expanding the amplitude equations in orders of the fluctuation potential, and truncating the doubles equations to first order in $U$. This is sufficient for the CC2 energy to be correct to second order in $U$.\cite{TheSecondOrderApproximateCC2Model}

The CC2 amplitude equations are given by\cite{TheSecondOrderApproximateCC2Model}
\begin{subequations}
\begin{equation}
    \Omega_{\mu_1} = \braM{\mu_1} \tilde{H} + [ \tilde{H}, T_2 ] \ketT{HF} = 0
\end{equation}
\begin{equation}
    \Omega_{\mu_2} = \braM{\mu_2} \tilde{H} + [ \tilde{F}, T_2 ] \ketT{HF} = 0,
\end{equation}
\end{subequations}
where the notation $\tilde{V}$ denotes T1-transformed operators, $\tilde{V} = \exp{-T_1} V \exp{T_1}$.

The CC2 Jacobian matrix
\begin{equation}
    \mathbf{A} = \begin{pmatrix}
        \mathbf{A}_{11} & \mathbf{A}_{12} \\
        \mathbf{A}_{21} & \mathbf{A}_{22} \\
    \end{pmatrix}
\end{equation}
is obtained by truncating the doubles component of the linear transformation with the Jacobian matrix $\mathbf{\rho} = \mathbf{Ar}$ to first order in $U$. This results in the Jacobian sub-blocks
\begin{equation} \label{eq:CC2_Jacobian}
\begin{aligned}
    A_{\mu_1\nu_1} &= \braM{\mu_1} \left[ \tilde{F}, \tau_{\nu_1} \right] 
    + \left[ \tilde{U}, \tau_{\nu_1} \right] 
    + \left[ \left[\tilde{U}, T_2\right], \tau_{\nu_1} \right] \ketT{HF} \\
    A_{\mu_1\nu_2} &= \braM{\mu_1} \left[ \tilde{U}, \tau_{\nu_2} \right] \ketT{HF} \\
    A_{\mu_2\nu_1} &= \braM{\mu_2} \left[ \tilde{U}, \tau_{\nu_1} \right] \ketT{HF} \\
    A_{\mu_2\nu_2} &= \braM{\mu_2} \left[ \tilde{F}, \tau_{\nu_2} \right] \ketT{HF} 
    = \epsilon_{\mu_2}\delta_{\mu_2\nu_2},
\end{aligned}
\end{equation}
where $\epsilon_\mu$ is the difference between the sum of MO energies of the excited determinant $\ketM{\mu}$ and the reference state. Finally, the CC2 ground state energy is obtained from the CCSD energy expression
\begin{equation}\label{eq:CCSD_energy}
    E_0 = \langle \text{HF} \vert \tilde{H} + [\tilde{H}, T_2] \vert \text{HF} \rangle,
\end{equation}
and the excitation energies are determined as the eigenvalues of the CC2 Jacobian.

\subsection{The SCC2 model}\label{sec:Theory_SCCTheory}

When truncating the standard coupled cluster expansion, matrix defects arise in the non-Hermitian coupled cluster Jacobian at near-degeneracies of same-symmetry excited states. At such conical intersections, the intersecting states collapse onto each other, and the intersection seam forms an $M-1$ dimensional intersection tube, where $M$ is the number of internal degrees of freedom of the system.\cite{CrossingConditionsInCoupledClusterTheory} The degeneracy is not lifted linearly in the branching plane ($gh$-plane)\cite{NonadiabaticityTheImportanceOfConicalIntersections} of the intersection, and inside the intersection tube, complex excitation energy pairs are encountered. \cite{CrossingConditionsInCoupledClusterTheory, ResolvingTheNotoriousCaseOfConicalIntersectionsForCoupledClusterDynamics} In contrast, in Hermitian methods,
no matrix defects appear in the Jacobian matrix. This guarantees real excitation energies and correct conical intersections seams of dimensionality $M-2$.\cite{UberDasVerhaltenVonEigenwertenBeiAdiabatischenProzessen, TheCrossingOfPotentialSurfaces, PotentialEnergySurfacesNearIntersections, NonadiabaticityTheImportanceOfConicalIntersections} SCC theory recovers these properties by enforcing linear independence of the intersecting states, which removes the matrix defects in the Jacobian matrix.\cite{ResolvingTheNotoriousCaseOfConicalIntersectionsForCoupledClusterDynamics} A brief overview of SCC theory will be given here. For a more extensive treatment of the theoretical framework, we refer to the literature.\cite{ResolvingTheNotoriousCaseOfConicalIntersectionsForCoupledClusterDynamics, AnOrbitalInvariantSimilarityConstrainedCoupledClusterModel, CoupledClusterTheoryForNonAdiabaticDynamics}

In SCC theory, a set of similarity constrained states is selected. These are the excited states for which linear independence is to be imposed. Throughout, we will assume two similarity constrained states, though it is in principle
possible to choose a larger number of constrained states.\cite{AnOrbitalInvariantSimilarityConstrainedCoupledClusterModel} The similarity constrained states, $\ketM{R^A}$ and $\ketM{R^B}$, are described as right EOM-CC excited states, with excited state amplitudes $\mathbf{r}^A$ and $\mathbf{r}^B$ 
and ground state contributions $r^A_0$ and $r^B_0$.

Linear independence of these states is imposed by enforcing orthogonality with respect to a positive semi-definite operator $P$:
\begin{equation} \label{eq:OrthogonalityCondition}
    O(A,B) = \braM{R^A} P \ketM{R^B} = 0.
\end{equation}
We will refer to this equation as the orthogonality condition. Several choices of $P$ have been explored.\cite{ResolvingTheNotoriousCaseOfConicalIntersectionsForCoupledClusterDynamics,AnOrbitalInvariantSimilarityConstrainedCoupledClusterModel, CoupledClusterTheoryForNonAdiabaticDynamics} In this work we apply the natural projection
\begin{equation} \label{eq:NaturalProjection}
    P = \sum_{\mu \ge 0} \ketM{\mu} \braM{\mu}.
\end{equation}

In order to enforce the orthogonality condition, additional flexibility in the wave function parameterization is needed. This is achieved by extending the standard cluster operator $T$ with an additional excitation operator $X$ scaled by the additional wave function parameter $\zeta$, 
\begin{equation}
    S = T + \zeta X.
\end{equation} 
From now on the superscript $S$ will refer to the SCC2 matrices and operators. In order to avoid redundancies in the cluster operator, $X$ must be linearly independent of $T$. The SCC cluster operator defines the SCC similarity transformed Hamiltonian according to eqn (\ref{eq:SimTransHamiltonian}).

The SCC ground state amplitudes are determined as in standard coupled cluster theory, see eqn (\ref{eq:AmplitudeEquations}). The excited state amplitudes of the similarity constrained states are determined as in EOM-CC theory, by solving the right eigenvalue equations, eqn (\ref{eq:UnsymmetricalEOMCC}). The ground and excited state equations are coupled via the orthogonality condition, eqn \eqref{eq:OrthogonalityCondition}. Therefore, the SCC equations, 
\begin{equation}\label{eq:SCC}
\begin{aligned}
    \mathbf{\Omega}^S = 0\\
    \mathbf{A}^S \mathbf{r}^A - \omega_A \mathbf{r}^A = 0\\
    \mathbf{A}^S \mathbf{r}^B - \omega_B \mathbf{r}^B = 0\\
    O(A,B) = 0,
\end{aligned}
\end{equation}
must be solved simultaneously. Other excited states can be determined from the SCC Jacobian with the SCC optimized ground state amplitudes and $\zeta$.

In our SCC2 model, we apply the same excitation operator $X_3$ used in the SCCSD model, \cite{AnOrbitalInvariantSimilarityConstrainedCoupledClusterModel}
\begin{equation}
     X = X_3 = \sum_{\mu_1\mu_2} \left( r^A_{\mu_1} r^B_{\mu_2} - r^B_{\mu_1} r^A_{\mu_2} \right) \tau_{\mu_1} \tau_{\mu_2}.
\end{equation}
In principle, other choices are possible, but there are strict requirements that $X$ should satisfy.\cite{AnOrbitalInvariantSimilarityConstrainedCoupledClusterModel} The choice of $X_3$ leads to the SCC2 cluster operator
\begin{equation}
    S = T_1 + T_2 + \zeta X_3.
\end{equation}

We expand the amplitude equations and the right Jacobian matrix eigenvalue equations in orders of the fluctuation potential $U$. As in CC2, the singles equations are treated exactly, whereas the doubles equations are treated to first order in the fluctuation potential. Since $r_{\mu_1}$ is treated as zeroth order in $U$, and $r_{\mu_2}$ is at least first order in the fluctuation potential, their product in $X_3$ is at least first order in $U$. We also treat $\zeta$ as zeroth order, such that $\zeta X_3$ is at least first order in $U$. As a result, the following amplitude equations are obtained:
\begin{equation}
\begin{aligned}
    \Omega_{\mu_1}^S &= \braM{\mu_1} \tilde{H} + [ \tilde{H}, T_2 ] + \zeta[ \tilde{H}, X_3 ]  \ketT{HF} = 0\\
    \Omega_{\mu_2}^S &= \braM{\mu_2} \tilde{H} + [ \tilde{F}, T_2 ] \ketT{HF}= 0.
\end{aligned}
\end{equation}
Note that the only SCC correction to the CC2 amplitude equations is the last term in the singles amplitude equations, whereas the doubles amplitude equations reduce to those of standard CC2. Further details about the derivation are given in the SI.

Since $X_3$ is a triple excitation, the SCC2 Jacobian matrix has no contributions from $X_3$ in the singles blocks, $\mathbf{A}_{11}$ and $\mathbf{A}_{12}$. In the doubles blocks, there is a second order SCC contribution. When using the CC2 definition of the Jacobian matrix, see eqn (\ref{eq:CC2_Jacobian}), the singles and doubles parts of the excited state equations become
\begin{equation}
\begin{aligned}
    \mathbf{A}_{11} \mathbf{r}_1 + \mathbf{A}_{12} \mathbf{r}_2 - \omega \mathbf{r}_1 &= 0\\
    \mathbf{A}_{21} \mathbf{r}_1 + \mathbf{A}_{22} \mathbf{r}_2 + \braM{\mu_2}[[\bar{U}, X_3], \tau_{\nu_1}] \ketT{HF} \mathbf{r}_1 - \omega \mathbf{r}_2 &= 0,
\end{aligned} \label{eq:SCC_Jacobi}
\end{equation}
where $\mathbf{r}$ is an excited state vector. However, following the truncations applied in CC2, the doubles part is treated up to first order in $U$, causing the term containing $X_3$ in eqn \eqref{eq:SCC_Jacobi} to be removed (as it is at least second order in $U$). 
Consequently, the SCC2 excited state equations fully reduce to their CC2 counterparts.

An expression for the overlap $O(A,B)$ using the natural projection at the singles and doubles level is available from the literature. \cite{AnOrbitalInvariantSimilarityConstrainedCoupledClusterModel} The ground state contributions to the similarity constrained states, $r^A_0$ and $r^B_0$, are calculated by eqn (\ref{eq:GroundStateContribution}). Here $X_3$ does not enter in $\mathbf{\eta}$, which is thus given by the standard coupled cluster expression, eqn (\ref{eq:eta}). Also the SCC2 ground state energy has the same expression as in CC2, given in eqn (\ref{eq:CCSD_energy}).

The SCC correction to the singles amplitude equations and the overlap equation are the only expressions of the SCC2 model not included in the standard CC2 model. As their determination has a computational scaling of $O(N^5)$ and $O(N^4)$, respectively, the SCC2 model maintains the  $O(N^5)$ computational scaling of CC2.

\section{Implementation}

The implementation of the SCC2 model reuses contributions from the existing implementation of the SCCSD method\cite{AnOrbitalInvariantSimilarityConstrainedCoupledClusterModel} 
in a development branch of the electronic structure program e$^T$.\cite{eT}

The CC2 contributions to the ground state amplitude equations $\mathbf{\Omega} = \mathbf{0}$, the transformation of a trial vector by the CC2 Jacobian matrix $\mathbf{A}\mathbf{r}$, the vector $\mathbf{\eta}$, and the CC2 ground state energy $E_0$ have previously been implemented in e$^T$.\cite{eT} An implementation-ready expression for the SCC2 correction term $\Omega^{\text{SCC}}_{\mu_1}$ is available in the original SCCSD paper,\cite{AnOrbitalInvariantSimilarityConstrainedCoupledClusterModel} as are expressions for the orthogonality condition using the natural projection. Among these expressions, those that distinguish the SCC2 model from CC2 are given in the SI.

In our SCC2 model, the amplitude equations, the EOM-CC right eigenvalue equations, and the orthogonality condition, are solved simultaneously using a direct inversion in the iterative subspace (DIIS)\cite{DIIS} algorithm, as detailed in Ref.\citenum{AnOrbitalInvariantSimilarityConstrainedCoupledClusterModel}. The initial guess for the wave function parameters can either be obtained from a preliminary CC2 calculation or from a previous SCC2 calculation from a neighboring molecular geometry. When using a previous SCC2 solution, the convergence is accelerated by diabatizing\cite{BAECK2003299} the 
MOs with respect to the previous geometry. If CC2 is used to initialize the calculation, the initial $\zeta$ is set to $0$.

A notable difference between SCC2 and SCCSD is the folding of the doubles amplitudes into the singles equations. In CC2, it is not necessary to solve the full set of the singles and doubles amplitude equations simultaneously. Instead, it is possible to calculate the doubles amplitudes as intermediates to be inserted into the singles amplitude equations.\cite{BeyondHertreeFock:MP2CC2} This reduces the dimensionality of the coupled set of equations to be solved, and thus the cost of each iteration. As the SCC2 model uses the same doubles amplitude equations as standard CC2, we implement the same folding, with only the singles amplitude equations being solved iteratively. The folding also allows a reduction in the memory requirements of SCC2 with respect to SCCSD. By calculating the doubles amplitudes on the fly, their storage can be avoided. 

\section{Results and Discussion}
\subsection{Conical Intersections}
\begin{figure}[bt!]
\includegraphics[width = 0.99\textwidth]{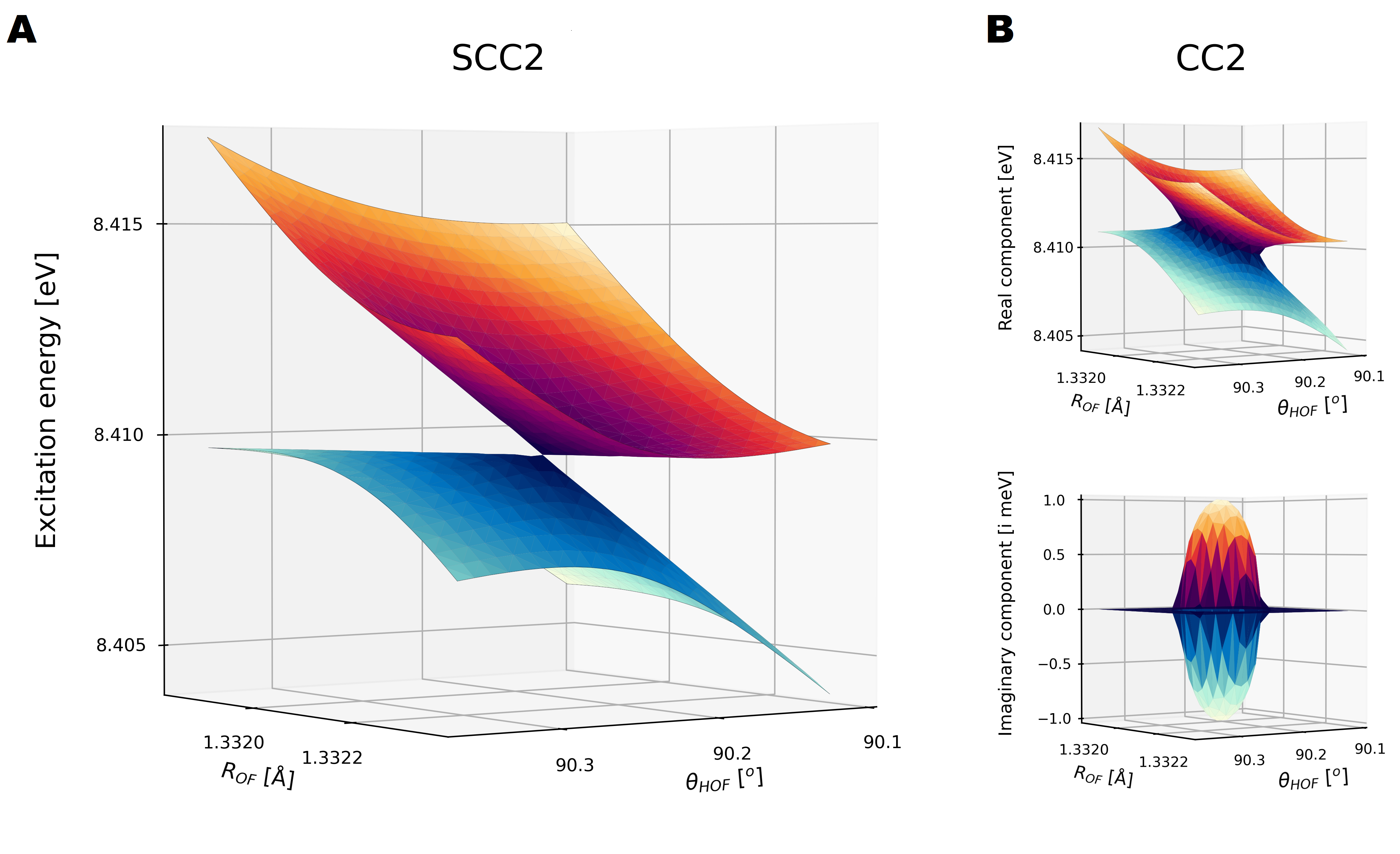}
\caption{\label{fig:HOF_CI} 
SCC2/aug-cc-pVDZ and CC2/aug-cc-pVDZ excitation energies at a conical intersection between the 1\textsuperscript{1}\textit{A}' and 2\textsuperscript{1}\textit{A}' states of HOF for an OH bond length of 1.1 Å. The calculations were executed with a residual threshold of $10^{-8}$ for an equidistant 25$\times$25 grid, extended with the explicit SCC2 intersection point, see the SI. (\textbf{A}) Real excitation energies obtained with the SCC2 model. With SCC2, the degeneracy is described as a point and is lifted linearly to first order. (\textbf{B}) Real and imaginary components of the CC2 complex excitation energies.}
\end{figure}

\begin{figure}[b!]
\includegraphics[width = 0.99\textwidth]{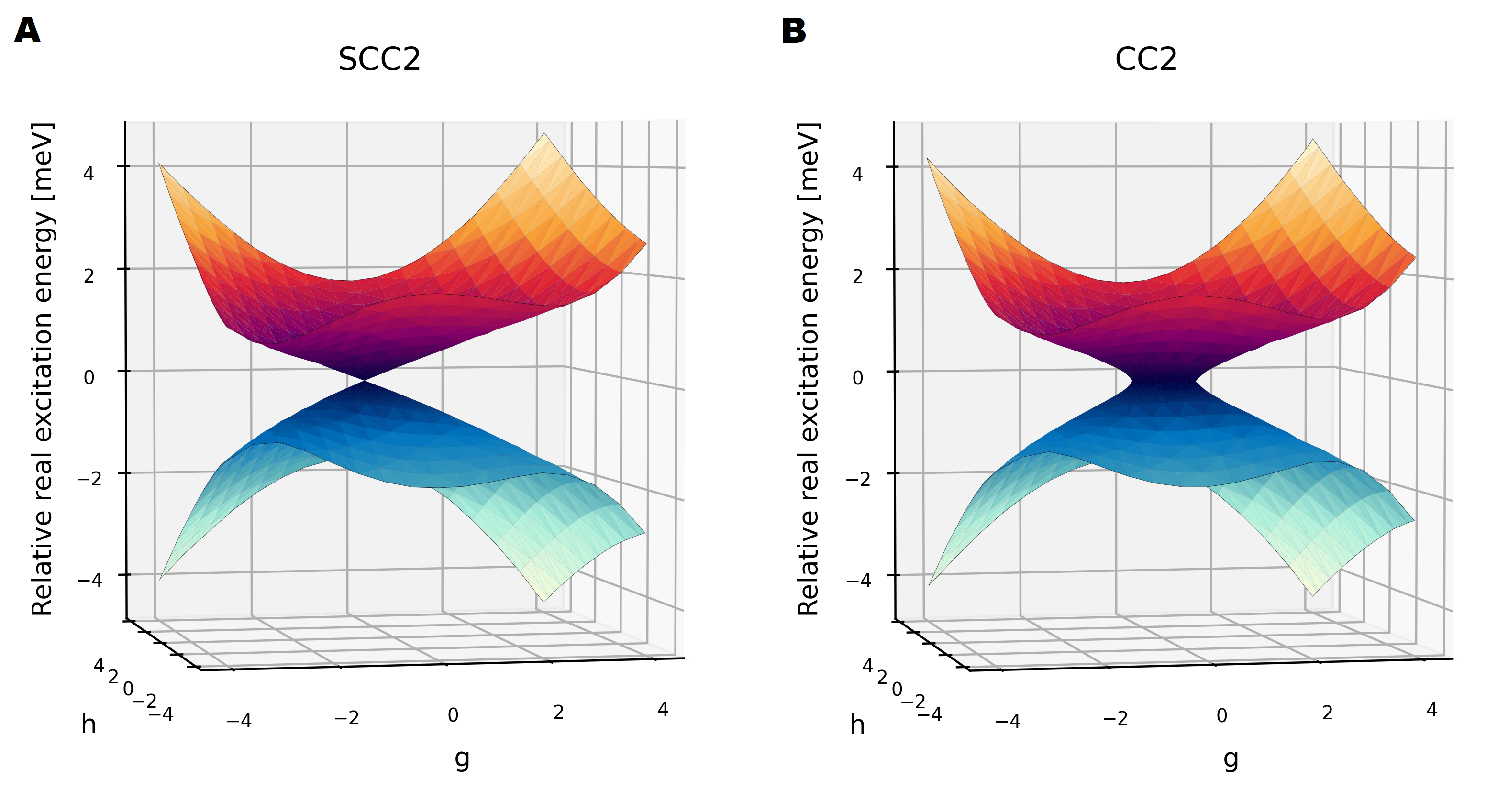}
\caption{\label{fig:Thymine_CI}  
SCC2/cc-pVDZ and CC2/cc-pVDZ excitation energies at a conical intersection between the $S_1$ and $S_2$ states in thymine. The excitation energies of the intersecting states are plotted relative to their average. We use an equidistant 17$\times$17 grid in the $gh$-plane, on $g,h \in [-4,4]$. This grid is extended with a tighter 19x19 grid in the area $g,h \in [-1,1]$. The SCC2 intersection point $(-0.6025, -0.0920)$ was included explicitly. Calculations were carried out with a residual threshold of $10^{-8}$. (\textbf{A}) SCC2 purely real relative excitation energies. (\textbf{B}) The real components of the CC2 relative excitation energies.}
\end{figure}

Hypofluorous acid (HOF) displays a conical intersection between the states $1^1A'$ and $2^1A'$. In Fig. \ref{fig:HOF_CI} we show this conical intersection modeled using SCC2 and CC2 in the subspace spanned by the OF bond length and the HOF angle, with the OH bond length kept fixed. The CC2 results display the expected unphysical artifacts of standard coupled cluster models. The intersection is an ellipse, corresponding to an erroneous $N-1$ dimensional intersection tube in internal coordinate space similar to that produced by CCSD.\cite{CrossingConditionsInCoupledClusterTheory} Moving away from the intersection ellipse, the degeneracy is not lifted linearly, and within the ellipse complex excitation energies are encountered. 

In contrast, the SCC2 model produces a correct conical intersection between the states. The intersection appears as a point in the scan, corresponding to an $N-2$ dimensional intersection seam within the full internal coordinate space. Moving away from the intersection, the degeneracy is lifted linearly, and no complex excitation energies are encountered. The SCC2 results are qualitatively similar to those obtained with SCCSD.\cite{AnOrbitalInvariantSimilarityConstrainedCoupledClusterModel} However, while SCCSD introduces SCC corrections in both the ground- and excited state equations, the SCC2 model corrects the unphysical artifacts of CC2 by only explicitly modifying the ground state singles amplitude equations. The SCC2 intersection geometry is reported in the SI.

To explore the behavior of the SCC2 model in a larger system, we consider the thymine molecule.
As a nucleobase, the photochemistry of thymine is of considerable interest,
and its decay mechanisms following photoexcitation by ultraviolet radiation has been studied extensively both theoretically and experimentally.\cite{unexpectedhydrogendissociationthymine, wolf2017probing, barbatti2015photoinduced, doi:10.1098/rsta.2017.0473,https://doi.org/10.1111/php.13903} In Fig. \ref{fig:Thymine_CI}, we show a $S_1$/$S_2$ conical intersection in thymine at the CC2 and SCC2 level, using $g$- and $h$-vectors obtained with SCCSD at a minimum energy conical intersection,\cite{CoupledClusterTheoryForNonAdiabaticDynamics} see the SI. Again, SCC2 corrects the unphysical artifacts produced by CC2 in the proximity of the conical intersection. The SCC2 intersection geometry is reported in the SI.

We note that changes in the excitation energies between SCC2 and CC2 in the results for HOF and thymine are in the range of few meV. This is two orders of magnitude below the typical CC2 error range,\cite{BenchmarksForElectronicallyExcitedStates:CASPT2CC2CCSDCC3, AMountaineeringStrategytoExcitedStates:HighlyAccurateEnergiesandBenchmarksforMediumSizedMolecules} indicating that the SCC2 model acts as a minor correction to the CC2 model.

\subsection{Computation Times}

\begin{table}[b!]
\centering
\caption{ CC2, SCC2, CCSD and SCCSD wall times for calculations on the ground state equilibrium geometry of adenine\cite{BenchmarksForElectronicallyExcitedStates:CASPT2CC2CCSDCC3} are provided with the number of iterations given in parentheses. Total times refer to the total wall time in the e$^T$ program. For the SCC calculations, the timings are reported for two different restart points, restarting from a CC solution and restarting from a SCC solution of a geometry 0.005Å in a random direction of internal coordinate space. The residual thresholds were set to $10^{-6}$. The CC2 and CCSD calculations were converged with a DIIS solver for the ground state amplitudes and a Davidson solver for the two lowest lying excited states. The calculations were executed on 48 cores on a dual-node Intel(R) Xeon(R) Gold 6342 system with 250 GB of memory}\label{tab:Adenine$^T$imings}
\begin{tabular}{l r r r}
Wall time [s]& CC2 & SCC2/CC2 & SCC2/rest.\\\hline
Ground state & (9) \, \, 1.89 & \multirow{ 2}{*}{(46) 59.38} & \multirow{ 2}{*}{(34) 43.15} \\
Excited state & (25) 16.06 &  & \\[2 pt]
Total & 24.06 & 86.27 & 47.62 \\
\multicolumn{3}{c}{}\\
Wall time [s]& CCSD & SCCSD/CCSD & SCCSD/rest.\\\hline
Ground state & (15) \, 37.65& \multirow{ 2}{*}{(55) 848.73} & \multirow{2}{*}{(31) 479.53} \\
Excited state & (24) 123.89 &  &  \\[2 pt]
Total & 167.60 & 1030.23 & 494.48 \\
\end{tabular}
\end{table}

\begin{figure*}[tb!]
\centering
\includegraphics[width = \textwidth]{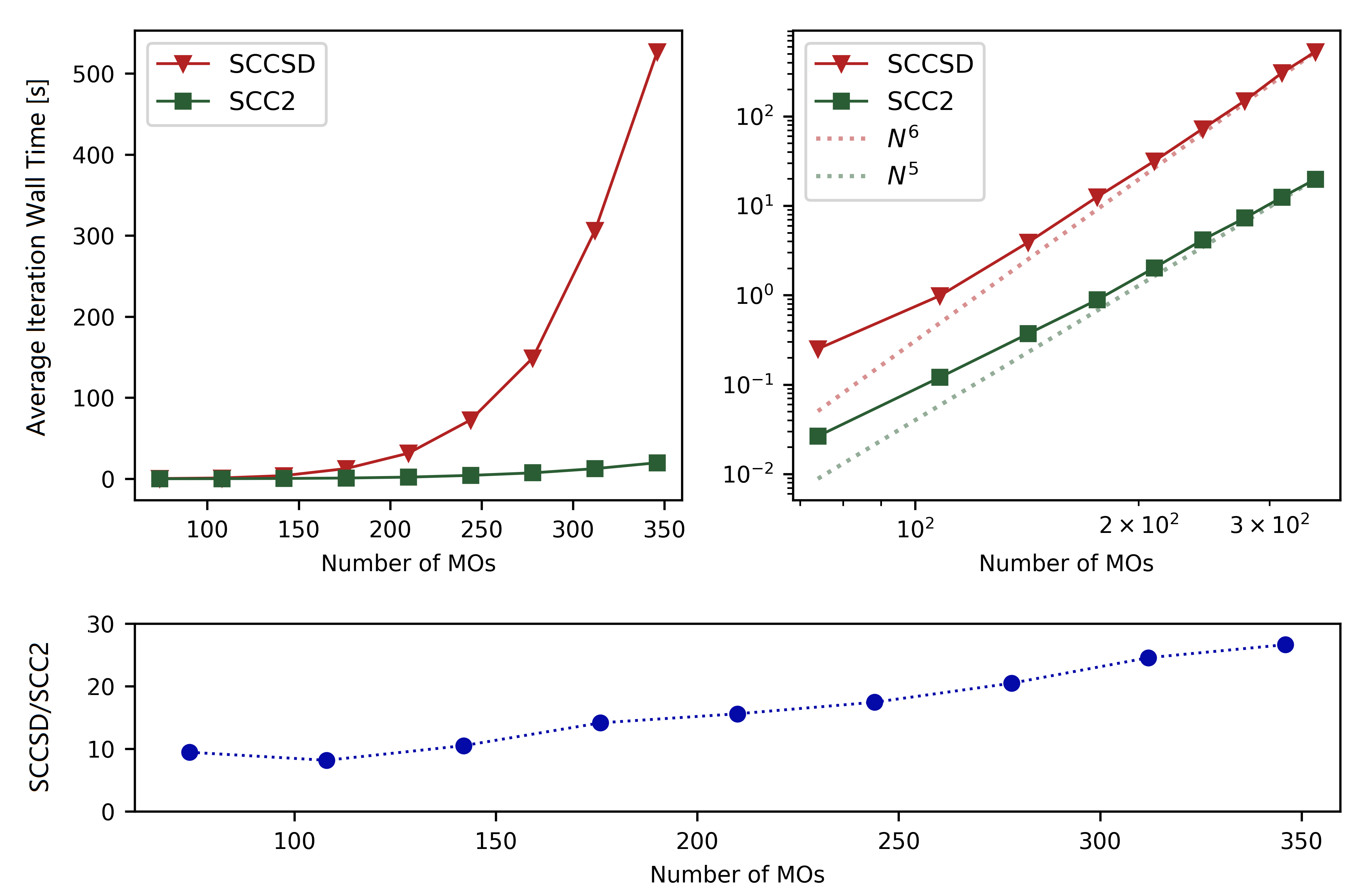}
\caption{\label{fig:Timings} 
SCC2/cc-pVTZ and SCCSD/cc-pVTZ calculation timings of the SCC DIIS solver for hypofluorous acid surrounded by 0-8 argon atoms situated at non-interacting distances ($>50$ Å) from the HOF molecule. For both methods, the 1\textsuperscript{1}\textit{A}' and 2\textsuperscript{1}\textit{A}' states of HOF 
were selected as the similarity constrained states and the HOF geometry is defined by $R_\text{OH} = 1.1 \text{Å}$, $R_\text{OF} = 1.33 \text{Å}$, $\theta_\text{HOF} = 90.5 ^\circ$ and is outside the intersection region. The $N^5$($N^6$) lines have slopes corresponding to the polynomial order, and intersect the data points of the largest SCC2(SCCSD) calculation. All calculations were converged to a residual threshold of $10^{-5}$. The calculations were run with 40 cores on an dual-node Intel(R) Xeon(R) Platinum 8480+ system with 2 TB of memory.}
\end{figure*}

In Table \ref{tab:Adenine$^T$imings} we report timings for SCC2 and SCCSD calculations on adenine at the ground state equilibrium geometry, together with the corresponding timings for CC2 and CCSD. For every SCC calculation, depending on the initial guess, two timings are reported. The SCC2 model is significantly faster than SCCSD in both cases, requiring less than one tenth of the time. Notably, when restarting the calculation from an SCC2 calculation at a neighboring geometry, the computation time was only about twice that of CC2. This indicates that SCC2 dynamics can be performed on a similar time frame as CC2 dynamics. 

Due to the difference in computational scaling between SCC2 and SCCSD, differences between their computation times are expected to increase with system size. To illustrate this aspect, we report timings for clusters of one HOF molecule surrounded by a varying number of argon atoms. The average iteration wall times for the determination of the solution of the SCCSD and SCC2 equations are reported in Fig. \ref{fig:Timings}, together with the ratios between calculation times of the two methods. For the largest system size, the SCC2 method is almost thirty times faster than SCCSD. An approximate linear trend in the ratio between SCCSD and SCC2 average iteration time is observed from 150 MOs. This is consistent with the theoretical scaling of the models.

\section{Conclusion and Perspectives}\label{sec:Conclusion}
We have developed an SCC2 model with the same computational scaling as the well-established CC2 model. As in CC2, SCC2 uses perturbation theory arguments to treat the double amplitudes to first order in the fluctuation potential. Results from initial calculations on same-symmetry conical intersections between excited states of hypofluorous acid and thymine show that the SCC2 model is able to correct the unphysical artifacts produced by CC2. The SCC2 model avoids complex excitation energies and produces correct intersection dimensionality and linearity. Beyond the intersection region, SCC2 introduces only minimal modifications to the CC2 potential energy surfaces.

Previously, we have encountered cases away from conical intersections where the SCC equations do not converge,\cite{CoupledClusterTheoryForNonAdiabaticDynamics} although such issues have so far not been observed in dynamics simulations. In our preliminary calculations, the SCC2 model was well-behaved in near-degenerate regions. Since the SCC correction is only needed in these regions, nonadiabatic dynamics simulations would likely not be affected by instabilities in other parts of the internal coordinate space when using an adaptive CC2/SCC2 algorithm in which SCC2 is only used to describe the conical intersection region. Such an adaptive algorithm for nonadiabatic dynamics requires handling possible non-negligible discontinuities in the potential energy surfaces when switching between the two different models, and is currently under development for CCSD/SCCSD. A successful application of this algorithm would be directly transferrable to the CC2/SCC2 case.

The use of SCC2 in nonadiabatic dynamics simulations is especially attractive as it constitutes a substantial reduction in computation times compared to SCCSD, which per now is the only other viable coupled cluster model for these applications. Specifically, our results indicate a 10--30 fold reduction in computation times for medium-sized systems. Furthermore, the lower computational scaling of the SCC2 model leads to larger reductions in computation time with increased system size. This makes it clear that SCC2 would enable coupled cluster nonadiabatic dynamics simulations on systems far larger than are practical at the SCCSD level. In order to run such simulations, the next step is the derivation and implementation of analytical energy gradients and derivative couplings for CC2 and SCC2. 

Conical intersections involving the ground state present additional challenges for single-reference methods beyond those associated with excited-state same-symmetry intersections. Recently, generalized coupled cluster theory (GCC)\cite{GCC} has resolved issues arising from the geometric phase effect\cite{berry1984quantal} when traversing ground state intersections,\cite{GPEinCC, kjønstad2024understandingfailureselectronicstructure} but the method does not avoid complex excitation energies at the ground state intersection. A hybrid model combining the properties of GCC and SCC is currently under development. Such a hybrid model could be used to describe photochemical pathways passing from higher excited states, through multiple conical intersections, and all the way back to the ground state.

\begin{acknowledgement}
We thank Marcus T. Lexander and Matteo Castagnola for discussions and assistance. L. S.,  S. A.. E. F. K. and H. K. acknowledge funding the European Research Council (ERC) under the European Union’s Horizon 2020 Research and Innovation Programme (Grant Agreement No. 101020016).
\end{acknowledgement}

\section*{Data availability}
Input and output files for calculations are available at zenodo.com at \\https://zenodo.org/records/15044176.

\section*{Author contributions}
H. K. and E. F. K. conceived the project. H. K. acquired funding and supervised the project. L. S. and E. F. K. implemented the method. L. S. ran calculations, with support from S. A.. L. S. wrote the first draft of the manuscript. All authors discussed the method and revised the manuscript.

\begin{suppinfo}
The supporting information includes derivations of the SCC2 amplitude equations, implementation ready expressions for relevant terms in the SCC2 equations, geometries for SCC2 conical intersections in HOF and thymine and g- and h- vectors used in the 2D-scan on thymine.
\end{suppinfo}

\bibliography{article}



\section*{Supplementary material}

# Supplementary Information

## Similarity Constrained CC2 for Efficient Coupled Cluster Nonadiabatic Dynamics

Leo Stoll, Sara Angelico, Eirik F. Kjønstad, and Henrik Koch*

*Department of Chemistry, Norwegian University of Science and Technology, Trondheim, Norway*

E-mail: henrik.koch@ntnu.no

# Contents

# 1 Derivations

We have defined the T1-transformed Hamiltonian $\bar{H}$ as[1]

$$\bar{H} = \exp(-\hat{T}_1)\hat{H}\exp(\hat{T}_1) = \bar{F} + \bar{U} \tag{1}$$

$$\bar{F} = \exp(-\hat{T}_1)\hat{F}\exp(\hat{T}_1) \tag{2}$$

$$\bar{U} = \exp(-\hat{T}_1)\hat{U}\exp(\hat{T}_1) \tag{3}$$

and apply the Fock-operators commutators with general excitation operators $\hat{\tau}_\mu, \hat{\tau}_\nu$ [1]

$$[\bar{F}, \hat{\tau}_\mu] = \epsilon_\mu \hat{\tau}_\mu \tag{4a}$$

$$[[\bar{F}, \hat{\tau}_\mu], \hat{\tau}_\nu] = 0 \tag{4b}$$

Then, using the SCC2 cluster operator $\hat{T} = \hat{T}_1 + \hat{T}_2 + \zeta \hat{X}_3$ where we consider $\hat{X}_3$ to be of first order in the fluctuation potential, and with the perturbative truncations outlined in the article, we obtain the following terms for the SCC2 amplitude equations:

$$\begin{aligned}
\Omega_{\mu_1} &= \langle \mu_1 | \exp(-\hat{T}_2 - \hat{X}_3)\, \bar{H}\, \exp(\hat{T}_2 + \hat{X}_3) | \text{HF} \rangle \\
&= \langle \mu_1 | \bar{H} + \underbrace{[\bar{F}, \hat{T}_2]}_{\text{operator rank}} + [\bar{U}, \hat{T}_2] + \underbrace{[\bar{F}, \hat{X}_3]}_{\text{operator rank}} + [\bar{U}, \hat{X}_3] | \text{HF} \rangle \\
&= \langle \mu_1 | \bar{H} + [\bar{U}, \hat{T}_2] + [\bar{U}, \hat{X}_3] | | \text{HF} \rangle
\end{aligned} \tag{5}$$

$$\begin{aligned}
\Omega_{\mu_2} &= \langle\mu_2|\exp(-\hat{T}_2 - \hat{X}_3)\,\bar{H}\exp(\hat{T}_2 + \hat{X}_3)|\text{HF}\rangle \\
&= \langle\mu_2|\bar{H} + [\bar{F}, \hat{T}_2] + \underbrace{\cancel{[\bar{U}, \hat{T}_2]}}_{\text{2}^{\text{nd}}\text{ order in }U} + \underbrace{\cancel{[\bar{F}, \hat{X}_3]}}_{\text{operator rank}} + \underbrace{\cancel{[\bar{U}, \hat{X}_3]}}_{\text{2}^{\text{nd}}\text{ order in }U} \\
&\quad + \underbrace{\tfrac{1}{2}[[\bar{F}, \hat{T}_2], \hat{T}_2]}_{Eq.(4)} + \underbrace{\tfrac{1}{2}\cancel{[[\bar{U}, \hat{T}_2], \hat{T}_2]}}_{\text{3}^{\text{rd}}\text{ order in }U} + \underbrace{\cancel{[[\bar{F}, \hat{T}_2], \hat{X}_3]}}_{Eq.(4)} + \underbrace{\cancel{[[\bar{U}, \hat{T}_2], \hat{X}_3]}}_{\substack{\text{operator rank} \\ \text{3}^{\text{rd}}\text{ order in }U}} \\
&\quad + \underbrace{\tfrac{1}{2}\cancel{[[\bar{F}, \hat{X}_3], \hat{X}_3]}}_{Eq.(4)} + \underbrace{\tfrac{1}{2}\cancel{[[\bar{U}, \hat{X}_3], \hat{X}_3]}}_{\substack{\text{operator rank} \\ \text{3}^{\text{rd}}\text{ order in }U}}|\text{HF}\rangle \\
&= \langle\mu_2|\bar{H} + [\bar{F}, \hat{T}_2]|\text{HF}\rangle.
\end{aligned} \qquad (6)$$

3

## 2 Implementation-ready Expressions for SCC2

The T1-transformed Hamiltonian of Eq. (1) can be expressed as the usual electronic Hamiltonian using the singlet operator, but with the T1-transformed one- and two electron integrals, respectively $\bar{h}_{pq}$ and $\bar{g}_{pqrs}$[1]

$$\begin{aligned}\bar{H} &= \sum_{pq} \bar{h}_{pq} E_{pq} \\ &+ \frac{1}{2} \sum_{pqrs} \bar{g}_{pqrs} \left( E_{pq} E_{rs} - \delta_{qr} E_{ps} \right) + h_{\text{nuc}}\end{aligned} \tag{7}$$

where $p, q, r, s \ldots$ are MO indices. Notably, the transformed two-electron integrals have reduced symmetry with respect to the non T1-transformed two-electron integrals, maintaining only particle symmetry $\bar{g}_{pqrs} = \bar{g}_{rspq}$.

Using the T1-transformed integrals, the SCC correction to the singles amplitude equations, $\Omega_{\mu_1}^{\text{SCC}}$, can be written on integral form as[2]

$$\Omega_{ai}^{\text{SCC}} = \sum_{bjck} \left( x_{ijk}^{abc} - x_{kji}^{abc} \right) \bar{L}_{jbkc} \tag{8}$$

with

$$x_{ijk}^{abc} = \zeta P^{-AB} \left( r_{ai}^A \tilde{r}_{bjck}^B + r_{bj}^A \tilde{r}_{aick}^B + r_{ck}^A \tilde{r}_{aibj}^B \right) \tag{9}$$

$$\bar{L}_{pqrs} = 2\bar{g}_{pqrs} - \bar{g}_{psrq} \tag{10}$$

$$\tilde{r}_{aibj}^k = r_{aibj}^k (1 + \delta_{ai,bj}) \tag{11}$$

$$P^{-AB} f(A, B) = f(A, B) - f(B, A). \tag{12}$$

As a right excited state is included as a bra in the orthogonality condition, due to the lack of orthonormality when using an adjoint bra basis,[1] the expression for the orthogonality

condition given in the article has to be extended with the elementary overlap matrix

$$S_{\mu\nu} = \langle \mu | \nu \rangle \tag{13}$$

and becomes[2]

$$\begin{aligned}\mathcal{O} =& R_0^A R_0^B (1 + \boldsymbol{q}^T \boldsymbol{S} \boldsymbol{q}) + R_0^A \boldsymbol{q}^T \boldsymbol{S} \boldsymbol{Q} \boldsymbol{R}^B \\ &+ (\boldsymbol{R}^A)^T \boldsymbol{Q}^T \boldsymbol{S} \boldsymbol{q} R_0^B + (\boldsymbol{R}^A)^T \boldsymbol{Q}^t \boldsymbol{S} \boldsymbol{Q} \boldsymbol{R}^B.\end{aligned} \tag{14}$$

The orthogonality condition can be calculated using the following expressions for $\boldsymbol{q}$

$$q_{ai} = t_i^a \qquad q_{aibj} = \frac{1}{1 + \delta_{ai,bj}} (t_{ij}^{ab} + t_i^a t_j^b) \tag{15}$$

and the following matrix transformations of a vector $\boldsymbol{c}$[2]

$$(\boldsymbol{Qc})_{ai} = c_{ai} \tag{16a}$$

$$(\boldsymbol{Qc})_{aibj} = c_{aibj} + \frac{1}{1 + \delta_{ai,bj}} (c_{ai} t_j^b + c_{bj} t_i^a) \tag{16b}$$

$$(\boldsymbol{c}^T \boldsymbol{Q}^T)_{ai} = c_{ai} + \sum_{ck} t_k^c b_{aick} \tag{17a}$$

$$(\boldsymbol{c}^T \boldsymbol{Q}^T)_{aibj} = c_{aibj} \tag{17b}$$

$$(\boldsymbol{Sc})_{ai} = 2 c_{ai} \tag{18a}$$

$$(\boldsymbol{Sc})_{aibj} = 2(1 + \delta_{ai,bj})(2 c_{aibj} - c_{ajbi}). \tag{18b}$$

# 3 CI Geometry for HOF

The conical intersection between the states $1^1A'$ and $2^1A'$ of HOF for an OH bond length of 1.1 Å was encountered at $R_{\text{OF}} = 1.3321938$Å and $\theta_{\text{HOF}} = 90.2475°$. This corresponds to the following geometry in cartesian coordinates:

Table 1: Geometry for SCC2/aug-cc-pVDZ conical intersection between the states $1^1A'$ and $2^1A'$ of HOF for an OH bond length of 1.1 Å. Geometry provided in Angstrom.

| Atom | $x$ | $y$ | $z$ |
|------|-----|-----|-----|
| H | -0.004750684189 | 1.099989741316 | 0.000000000000 |
| O | 0.000000000000 | 0.000000000000 | 0.000000000000 |
| F | 2.517481426844 | 0.000000000000 | 0.000000000000 |

# 4 Geometries for 2D-scan of Thymine

The basis vectors used to span the $gh$-plane used in the 2D-scan of thymine were obtained from the $g$- and $h$-vectors available from SCCSD calculations of a minimum energy conical intersection (MECI) in a distorted (non-planar) geometry.[3] The geometry for the $S_1/S_2$ SCC2 conical intersection was located within the branching plane of this SCCSD MECI.

Table 2: Geometry for SCC2/cc-pVDZ conical intersection between states $S_1$ and $S_2$ located in thymine 2D-scans. Geometry provided in bohr.

| Atom | $x$ | $y$ | $z$ |
|---|---|---|---|
| H | 3.595889643580 | 3.282432604151 | 0.975922146276 |
| H | -1.154079844480 | 4.850251870840 | -0.995000020486 |
| H | 2.184193690383 | -3.852070634671 | -0.143708705776 |
| H | -5.183882375697 | 3.738832874934 | 1.242990457532 |
| H | -6.200931791418 | 0.101331649512 | 0.678594175727 |
| H | -5.493251220592 | 2.111726044220 | -1.734813526935 |
| C | 3.071302454392 | -0.159842689030 | -0.013548339899 |
| C | -0.403787069982 | 3.023873528098 | -0.362747000370 |
| C | -1.529548654373 | -1.596338517454 | -0.154071584974 |
| C | -2.203312576980 | 1.062221681888 | -0.012296553478 |
| C | -4.981786031830 | 1.811395162821 | 0.241722276629 |
| N | 1.186491679583 | -1.904180617241 | -0.153756142661 |
| N | 2.083083366355 | 2.288810152087 | 0.107079474525 |
| O | 5.307053702028 | -0.536037647085 | 0.107398024976 |
| O | -2.928617287789 | -3.497304711432 | 0.108434295250 |

To obtain a suitable depiction of the conical intersection, first the $h$-vector was orthogonalized with respect to the $g$-vector. Then both vectors were normalized, upon which they were rotated counterclockwise by three degrees in the plane they span, $v = \cos 3° g + \sin 3° h$ and $w = -\sin 3° g + \cos 3° h$, where $v$ corresponds to $g$ and $w$ corresponds to $h$ as labeled in the article. Finally, $v$ and $w$ were rescaled as to produce a nearly circular conical intersection in CC2/cc-pVDZ calculations. The final vectors $v$ and $w$ are given below.

Table 3: $v$-vector used to span the $gh$-plane in thymine 2D-scans.

| Atom | $x$ | $y$ | $z$ |
|------|-----|-----|-----|
| H | -0.00910526 | -0.01966916 | -0.00470308 |
| H | -0.01248047 | 0.00582823 | -0.00378694 |
| H | -0.0169608 | 0.05170705 | 0.00113383 |
| H | 0.00284974 | 0.01112344 | -0.00131754 |
| H | -0.01329086 | -0.02306126 | -0.00640475 |
| H | 0.00267549 | -0.00649999 | -0.00127202 |
| C | -0.01362692 | -0.01236836 | -0.00871645 |
| C | 0.00787804 | -0.00810476 | -0.0044191 |
| C | -0.01722977 | 0.01494915 | -0.00565207 |
| C | -0.00594766 | 0.00842545 | -0.00216768 |
| C | 0.01153683 | 0.02004446 | 0.01598482 |
| N | 0.05726479 | -0.05634807 | 0.00546124 |
| N | 0.00775673 | 0.02887218 | 0.01335076 |
| O | 0.00602834 | -0.00351277 | 0.00137911 |
| O | -0.0073482 | -0.01138553 | 0.00112988 |

Table 4: $w$-vector used to span the $gh$-plane in thymine 2D-scans.

| Atom | $x$ | $y$ | $z$ |
|------|-----|-----|-----|
| H | -1.64401523e-04 | -4.75539068e-05 | -5.23155035e-05 |
| H | -6.29998517e-05 | -6.42700521e-05 | 1.10981744e-04 |
| H | 1.89669016e-05 | -9.62072649e-05 | 5.21042299e-06 |
| H | 8.00721628e-06 | -9.03302695e-05 | 2.73396897e-05 |
| H | -4.74427001e-05 | 8.57998386e-05 | 2.20447142e-05 |
| H | -1.01652755e-04 | 6.71906680e-06 | 8.44935763e-05 |
| C | 1.95697650e-04 | 1.46616735e-03 | 9.33302245e-05 |
| C | -8.25662194e-04 | -7.64532457e-04 | 3.77165417e-04 |
| C | 1.39151644e-03 | 2.23063732e-03 | -2.88562194e-04 |
| C | 3.22753138e-04 | 3.65827178e-04 | -5.12915698e-05 |
| C | -2.16299607e-04 | -2.52840985e-06 | -4.05688216e-04 |
| N | -4.13791871e-04 | -6.65569635e-04 | 3.46923635e-05 |
| N | 8.83866255e-04 | -5.10535001e-04 | -2.51239393e-04 |
| O | -2.11635332e-04 | -4.33079509e-04 | -2.25471634e-05 |
| O | -7.76911146e-04 | -1.48050989e-03 | 3.16392145e-04 |



\end{document}